\documentstyle[sprocl,epsfig]{article}

\def\st{\scriptstyle}

\def\ra{\rightarrow}

\def\be{\begin{equation}}
\def\ee{\end{equation}}
\def\bea{\begin{eqnarray}}
\def\eea{\end{eqnarray}}

\def\st{\mbox{$\rm \tilde{t}_1$}}
\def\s2{\mbox{$\rm \tilde{t}_2$}}
\def\sl{\mbox{$\rm \tilde{t}_L$}}
\def\sr{\mbox{$\rm \tilde{t}_R$}}
\def\mt{\mbox{$m_{\rm \tilde{t}_1}$}}
\def\m2{\mbox{$m_{\rm \tilde{t}_2}$}}
\def\cost{\mbox{$\cos\theta_{\rm \tilde t}$}}
\def\sint{\mbox{$\sin\theta_{\rm \tilde t}$}}
\def\ee{\mbox{$\rm e^+e^-$}}

\begin{document}

\topmargin = 0cm

{

\begin{titlepage}

\begin{center}
\mbox{ } 

\vspace*{-4cm}


\end{center}
\vskip 0.5cm
\begin{flushright}
\Large
\mbox{\hspace{10.2cm} hep-ph/9911345} \\
\mbox{\hspace{10.2cm} IEKP-KA/99-20} \\
\mbox{\hspace{11.5cm} Nov. 1999}
\end{flushright}
\Large
\begin{center}
\vskip 3cm
{\Large\bf
 STUDY OF SCALAR TOP QUARKS AT \\
 A FUTURE \boldmath$\rm e^+e^-$\unboldmath\ LINEAR COLLIDER}
\vskip 1.2cm
{\Large\bf 
M. Berggren$^a$, 
R. Ker\"anen$^b$,\\
H. Nowak$^c$,
A. Sopczak$^b$}

{\Large
\vspace*{2cm}
$^a$~LPNHE, Universit\'e de Paris VI \& VII

\vspace*{0.5cm}
$^b$~Karlsruhe University

\vspace*{0.5cm}
$^c$~DESY Zeuthen }
\end{center}

\vskip 1.8cm
\centerline{\Large \bf Abstract}

\vspace*{2cm}
\hspace*{-3cm}
\begin{picture}(0.001,0.001)(0,0)
\put(,0){
\begin{minipage}{16cm}
\Large
\renewcommand{\baselinestretch} {1.2}
The scalar top discovery potential has been studied
with a full-statistics background simulation for $\sqrt{s}=500$~GeV
and ${\cal L}=500$~fb$^{-1}$. The simulation is based on a fast and
realistic simulation of a TESLA detector. The large simulated data 
sample allowed the application of an Iterative Discriminant Analysis 
(IDA) which led to a significantly higher sensitivity than in 
previous studies. The effects of beam polarization on signal efficiency
and individual background channels are studied using separate optimization
with the IDA for both polarization states. The beam polarization is very 
important to measure the scalar top mixing angle and to determine its 
mass. Simulating a 180 GeV scalar top at minimum production
cross section, we obtain $\Delta m=1$~GeV and $\Delta\cost=0.009$.
\renewcommand{\baselinestretch} {1.}

\normalsize 
\vspace{0.5cm}
\begin{center}
{\large \em
Talk at the Worldwide Workshop on Future $e^+e^-$ Collider, 
April 1999, Sitges, Spain, \\
to be published in the proceedings.
\vspace*{-6.5cm}
}
\end{center}
\end{minipage}
}
\end{picture}
\vfill

\end{titlepage}

\thispagestyle{empty}
\mbox{ }
\newpage
\setcounter{page}{1}
}

\title{STUDY OF SCALAR TOP QUARKS \\
       AT A FUTURE \boldmath$\rm e^+e^-$\unboldmath\ LINEAR COLLIDER}

\author{R. KERANEN, A. SOPCZAK\footnote{speaker}}

\address{Karlsruhe University}

\author{H. NOWAK}

\address{DESY-Zeuthen}

\author{M. BERGGREN}

\address{LPNHE, Universit\'e de Paris VI \& VII}


\maketitle\abstracts{
\vspace*{1cm}
The scalar top discovery potential has been studied
with a full-statistics background simulation for $\sqrt{s}=500$~GeV
and ${\cal L}=500$~fb$^{-1}$. The simulation is based on a fast and
realistic simulation of a TESLA detector. The large simulated data 
sample allowed the application of an Iterative Discriminant Analysis 
(IDA) which led to a significantly higher sensitivity than in 
previous studies. The effects of beam polarization on signal efficiency
and individual background channels are studied using separate optimization
with the IDA for both polarization states. The beam polarization is very 
important to measure the scalar top mixing angle and to determine its 
mass. Simulating a 180 GeV scalar top at minimum production
cross section, we obtain $\Delta m=1$~GeV and $\Delta\cost=0.009$.}

\vspace*{1cm}
\section{Introduction}
The study of the scalar top quarks is of particular interest, since the
lighter stop mass eigenstate is likely to be the lightest scalar quark in a 
supersymmetric theory. The mass eigenstates are \mt\ and \m2\ with 
$\mt < \m2$, where $\st=\cost \sl + \sint \sr$ and 
$\s2=-\sint \sl + \cost \sr$. We study the experimental possibilities to
determine \mt\ and \cost\ at a high-luminosity \ee\ linear collider 
like the TESLA project~\cite{tesla} with the possibility of polarizing 
the $\rm e^-$ beam.
The MSSM cross section was calculated with CALVIN2.0~\cite{xsec}.

Previously, the discovery potential for scalar top quarks was simulated
for $\sqrt{s}=500$~GeV and ${\cal L}=10$~fb$^{-1}$ where sequential
cuts for the event selection were applied~\cite{morioka,moriokaproc,desy123d}.
The possibility of beam polarization to determine mass and mixing angle was 
studied~\cite{munich,desy123e} and resulting estimates of the errors of the 
soft-breaking parameters in a supersymmetric theory were given~\cite{zphys}. 
The 10~fb$^{-1}$ study gave $4.3\%$ signal efficiency with 
21 signal and 9 background events, 
which resulted in $\Delta m=7$~GeV and $\Delta\cost=0.06$.

\clearpage
\section{Event Simulation}
The simulated production process is $\ee\ra\st\bar{\st}$, where subsequently
each scalar top decays into a c-quark and a neutralino 
which escapes detection.
The resulting signature is two jets and large missing energy.
This channel is dominant unless the decay into a chargino is 
kinematically allowed. The previous 10~fb$^{-1}$ analysis gave 
slightly higher sensitivity in the chargino channel.
The signal generator~\cite{asgenerator} includes initial
state radiation and beamstrahlung. The generated 
events are passed through
the parametric detector simulation SGV~\cite{sgv} tuned for a 
TESLA detector~\cite{tesla}. We simulate a 180 GeV scalar top 
and a 100 GeV neutralino.

\section{Event Preselection}
The event selection consists of several steps. First, an event preselection
is applied using the hadronic character and large missing energy of the 
simulated signal. 
The same preselection as for the 10~fb$^{-1}$ study is used and we checked 
that similar fractions of events in each background channel pass in this
500~fb$^{-1}$ analysis. 
The following preselection cuts are applied:
$25 < N_{\rm cluster} < 110$, 
$0.2 < E_{\rm vis} / \sqrt{s} < 0.7$,
$E^{\rm imbalance}_{\parallel} / E_{\rm vis} < 0.5$,
thrust  $ < 0.95$,
$|\cos\theta_{\rm thrust}| < 0.7$.
The number of simulated events for the
signal and for each background channel, as well as the remaining events
after the preselection, are given in Table~\ref{tab:pre}.
\begin{table}[hp]
\vspace*{-0.3cm}
\caption{\label{tab:pre} Number of simulated signal and background events before and 
                         after the preselection.}
\begin{center}
\begin{tabular}{|c|c|c|c|c|c|c|c|}\hline
Channel      & $\tilde{\chi}^0$c$\tilde{\chi}^0\mathrm{\bar{c}}$ &$\rm q\bar{q}$  & WW   &eW$\nu$& 
t$\rm \bar{t}$&ZZ   & eeZ   \\ \hline
(in 1000)    & 50  &6250 & 3500 & 2500  & 350 & 300 & 3000 \\ \hline
After presel. & 47\% &46788 &115243&252189&43759&4027 & 4069  \\ \hline
\end{tabular}
\end{center}
\vspace*{-0.7cm}
\end{table}

\section{Iterative Discriminant Analysis}
In order to separate the signal from the background, 
the following selection variables are defined: 
visible energy,
number of jets, 
thrust value and direction, 
number of clusters, 
transverse and parallel imbalance, 
acoplanarity and invariant mass of two formed jets.

\begin{figure}
\vspace*{-0.4cm}
\caption{\label{fig:eseen} Visible energy and transverse momentum after the preselection.}
\begin{center}
\vspace*{-0.7cm}
\mbox{\epsfig{file=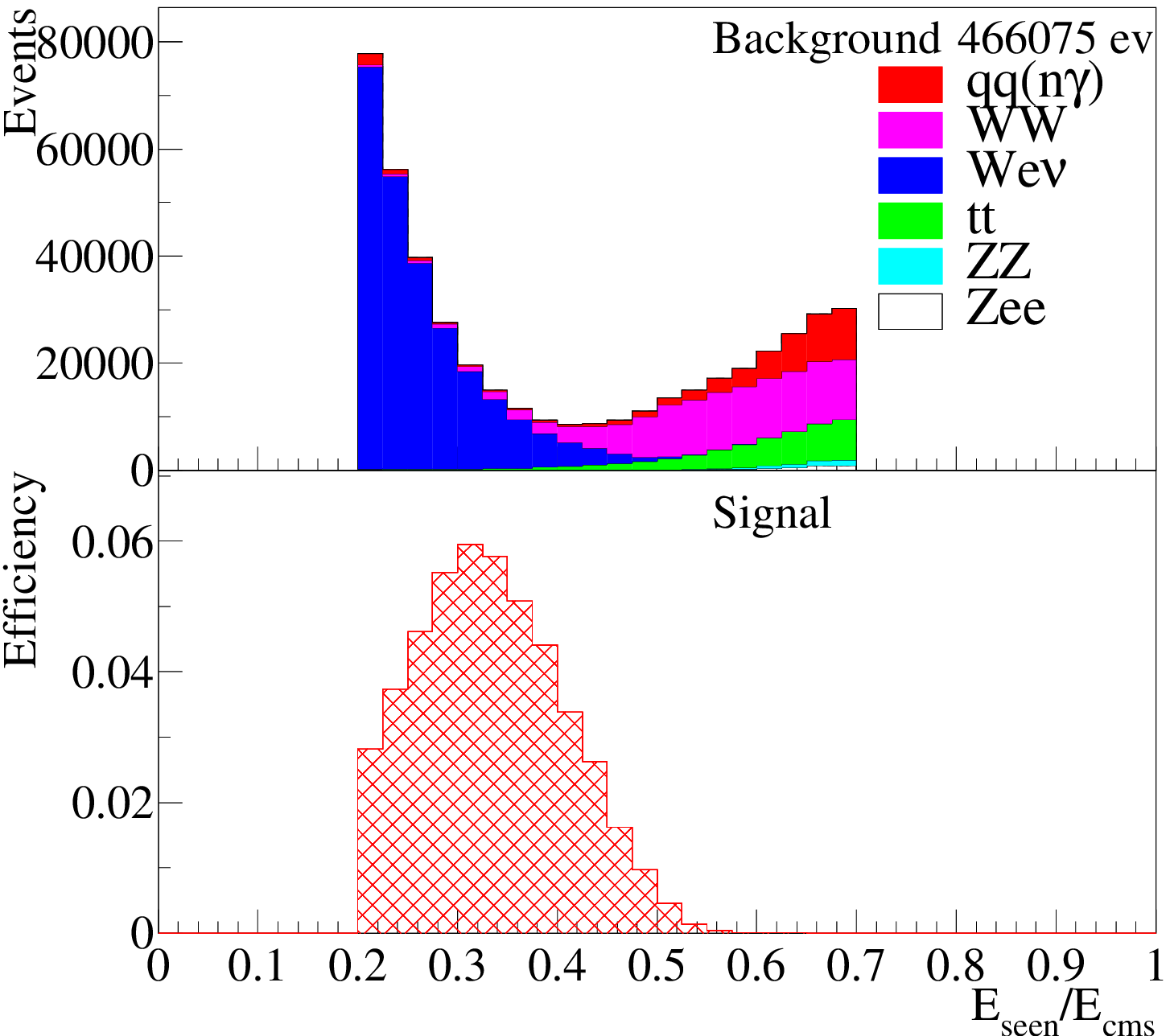,width=0.49\textwidth}}
\mbox{\epsfig{file=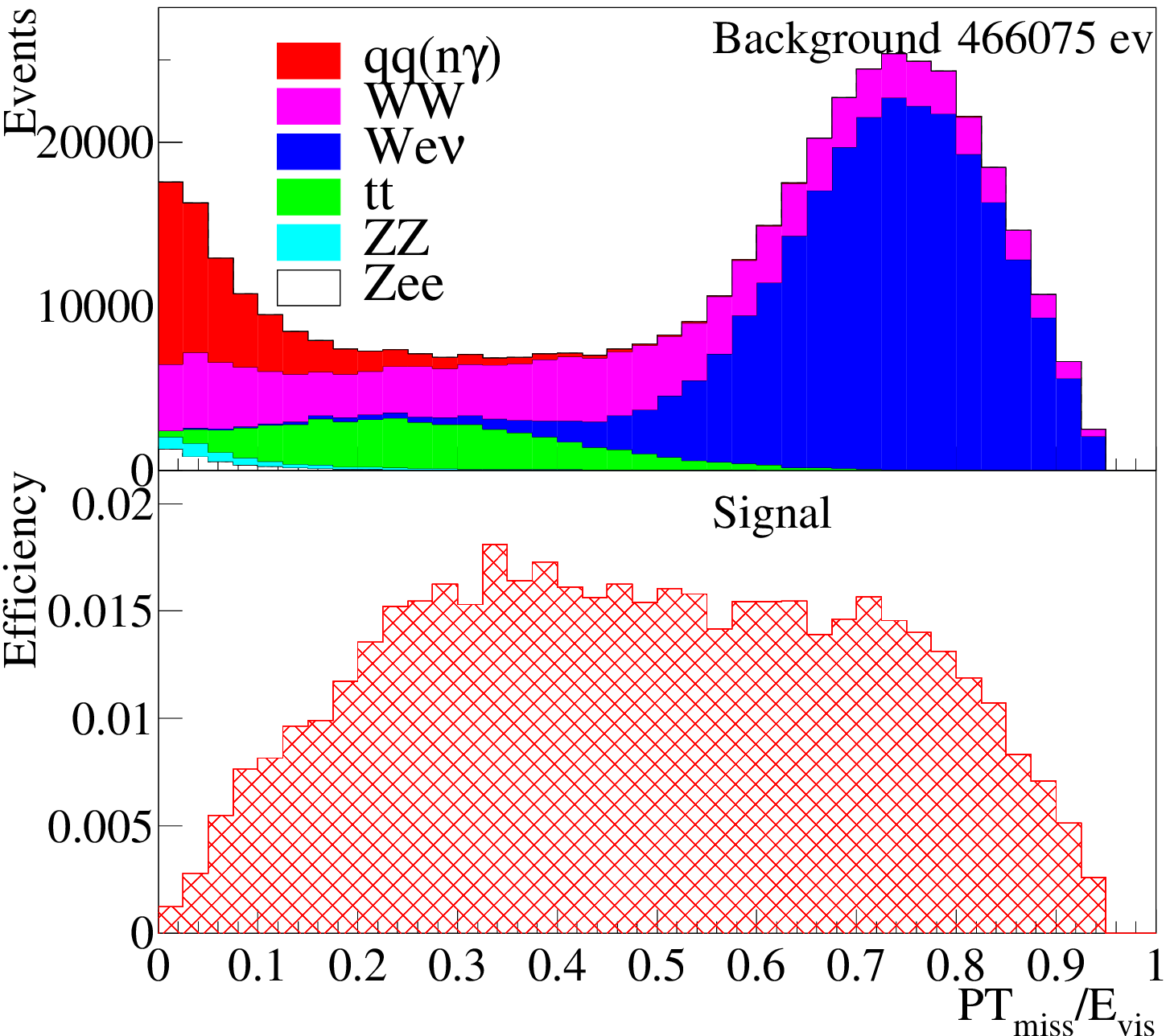,width=0.49\textwidth}}
\end{center}
\vspace*{-0.5cm}
\end{figure}

\begin{figure}
\caption{\label{fig:ida1} First step IDA output and resulting thrust
values.}
\begin{center}
\vspace*{-0.7cm}
\mbox{\epsfig{file=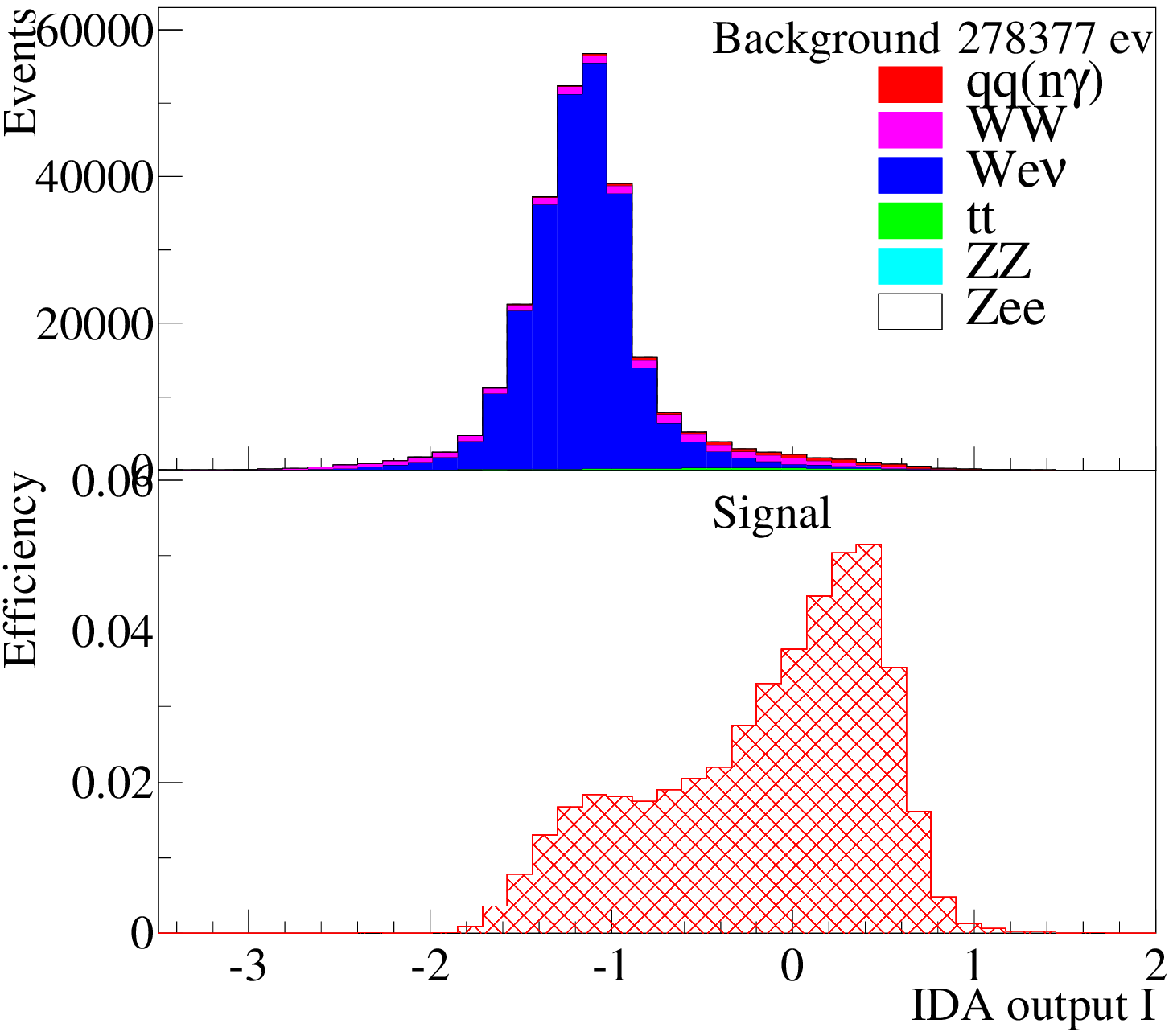,width=0.49\textwidth}}
\mbox{\epsfig{file=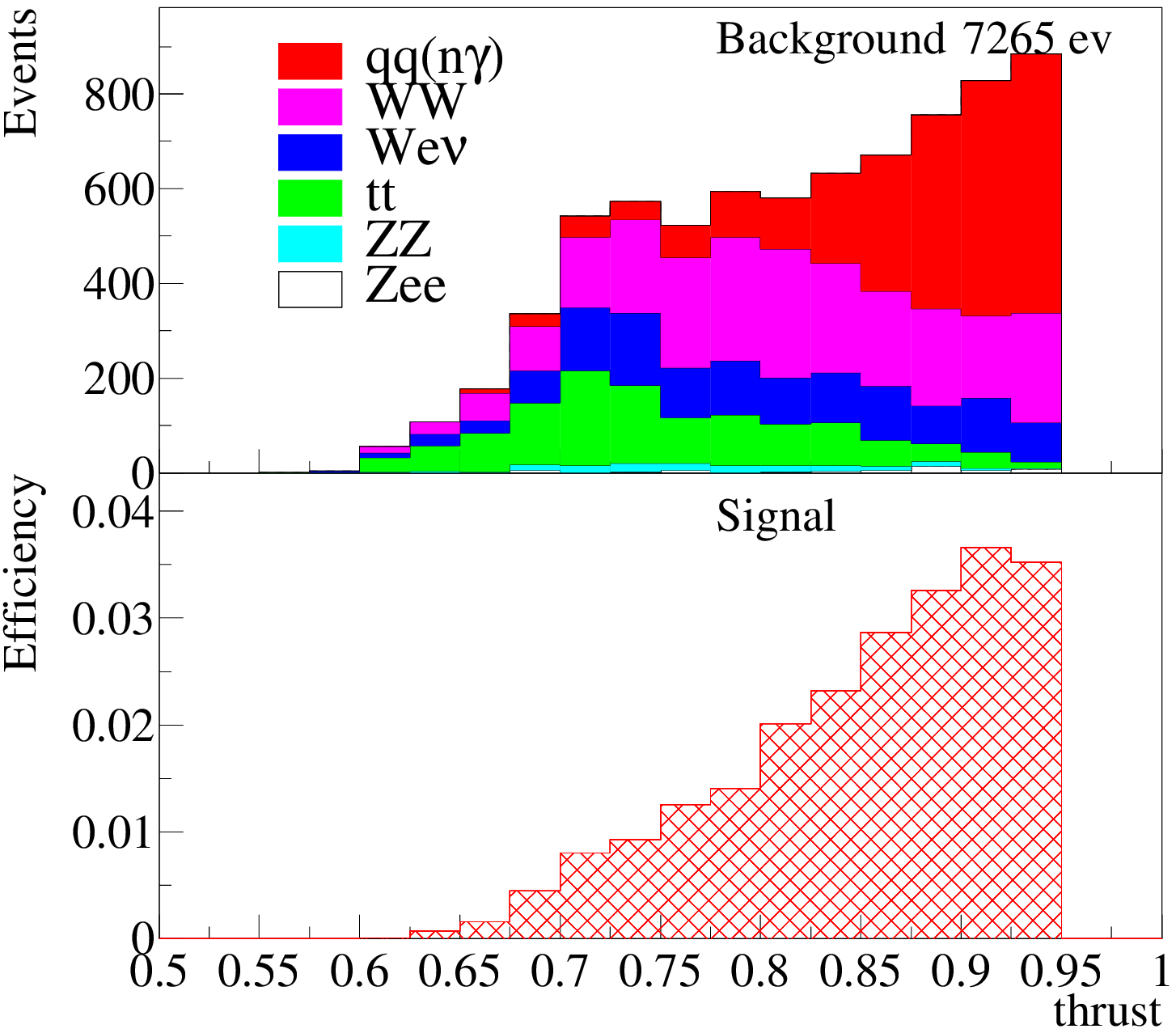,width=0.49\textwidth}}
\end{center}
\vspace*{-0.5cm}
\end{figure}

\begin{figure}
\caption{\label{fig:ida2} Final IDA output and background vs. signal 
efficiency 
for a 180 GeV scalar top and a 100 GeV neutralino.}
\begin{center}
\vspace*{-0.7cm}
\mbox{\epsfig{file=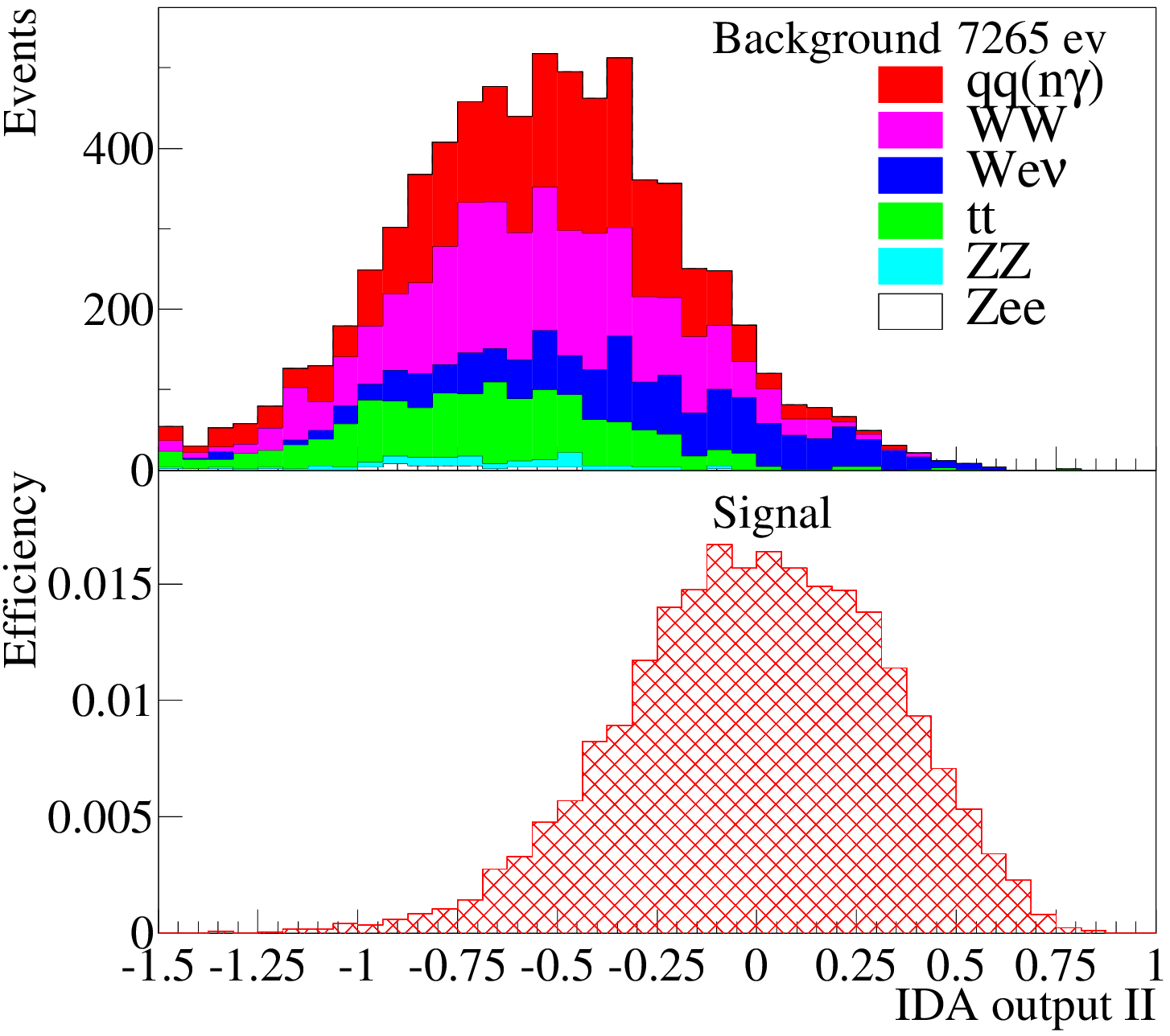,width=0.49\textwidth}}
\mbox{\epsfig{file=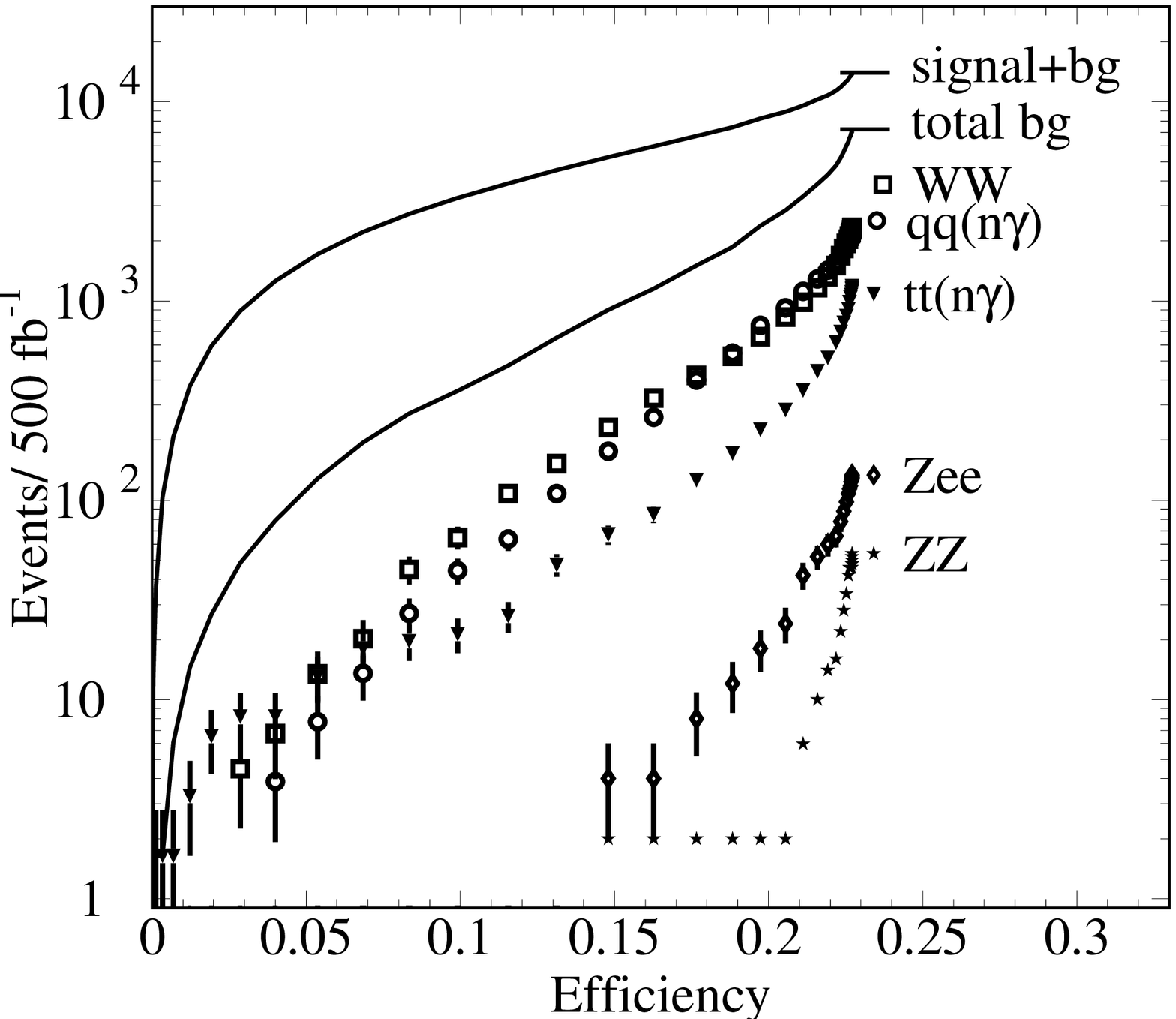,width=0.49\textwidth}}
\end{center}
\vspace*{-0.5cm}
\end{figure}

Figure~\ref{fig:eseen} shows the
simulated visible energy and transverse momentum after the preselection.
Following a tighter preselection,
$E_{\rm vis} / \sqrt{s} < 0.52$ and
$N_{\rm cluster} < 80$,
278377 background events remain.
Half of these events and half of the signal events are used to train 
the IDA~\cite{ida}.
In a first step, a cut on the IDA output variable is applied,
defined by a reduction in the signal of 50\%.
The IDA output variable and the thrust value for
the remaining signal and 7265 background events are shown 
in Fig.~\ref{fig:ida1}.
These events are  again passed through the IDA. Figure~\ref{fig:ida2} shows the
IDA output variable and the resulting number of background events as a 
function of the signal efficiency. 
For 12\% efficiency, 400 background events are expected.

\section{Effects of Beam Polarization}
\begin{figure}[hp]
\caption{\label{fig:pol} Background vs. signal efficiency 
for left- and right-polarization.}
\vspace*{-0.7cm}
\begin{center}
\mbox{\epsfig{file=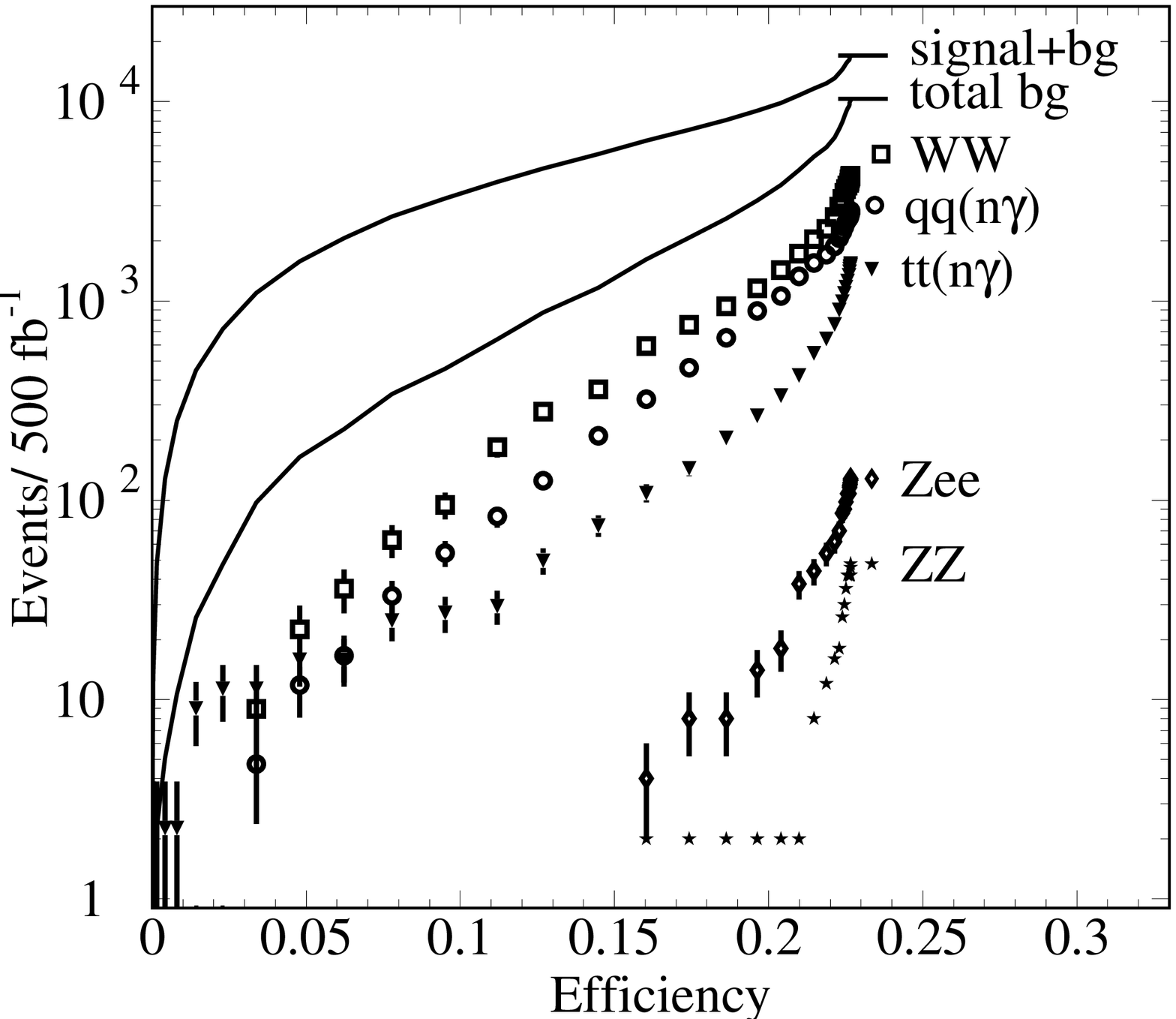,width=0.49\textwidth}}
\mbox{\epsfig{file=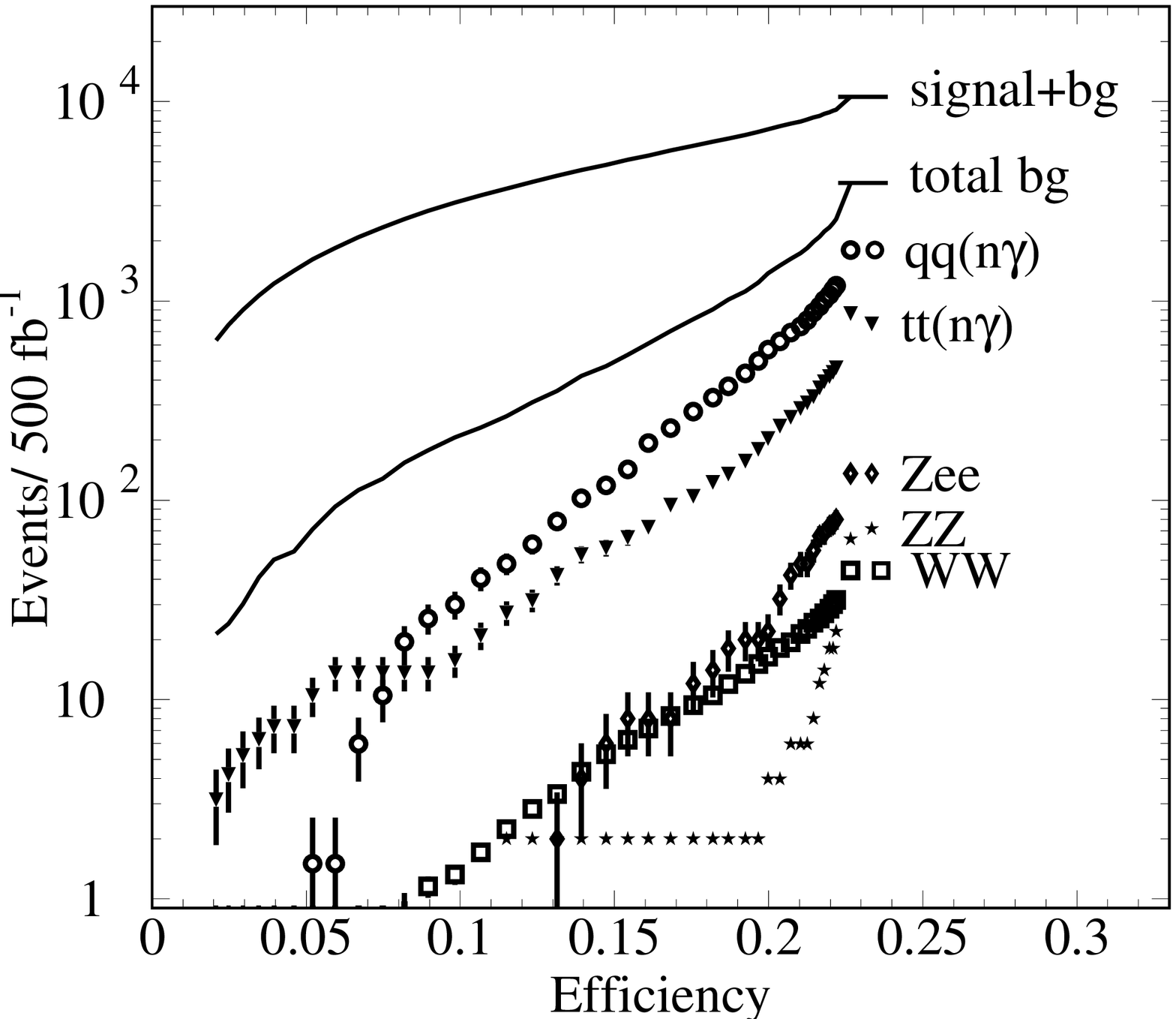,width=0.49\textwidth}}
\end{center}
\vspace*{-0.5cm}
\end{figure}

The polarization of the $\rm e^-$ beam at a future linear collider
offers the opportunity to enhance and suppress the left- and right-handed 
couplings of the scalar top signal and to determine mass and mixing angle 
independently. The production cross section of each background process
depends differently on the polarization. It is therefore important for a 
high-statistics analysis to study the expected background channels 
individually.
The expected cross sections are given in 
Table~\ref{tab:bgxsec}~\cite{generator}.
The IDA analysis is repeated for $-0.9$ and $0.9$ 
polarization~\footnote{For a polarization of 
$-0.9$, 95\% of the $\rm e^-$ are left-polarized.
In the previous analyses~\cite{munich,desy123e,zphys}
it was assumed that only 90\% of the $\rm e^-$ were polarized.}
in order to take into account the different composition 
of the background and the expected signal cross sections.
Figure~\ref{fig:pol} shows the number of background events
as a function of the signal efficiency for left- and right-polarization.
For 12\% detection efficiency, 650 background events are expected
leading to $\sigma_{\rm left} = 54.5 \pm 1.0 $ fb, 
and 240 background events giving $\sigma_{\rm right} = 50.9 \pm 1.0 $ fb, 
where $\Delta\sigma/ \sigma = 
\sqrt{N_{\rm signal} + N_{\rm background}} / N_{\rm signal}$.

\begin{table}[hp]
\vspace*{-0.2cm}
\caption{\label{tab:bgxsec} Background cross sections (pb) from different 
                            event generators for $\rm e^-$ polarization.}
\begin{center}
\begin{tabular}{|c||c|c|c|c|c|c|} \hline
 Pol.         &$\rm W e \nu$ &  WW & $\rm q\bar q$ &$\rm t\bar t$ &  ZZ & Zee  \\
 of $\rm e^-$     
&{\small \hspace*{-0.5mm}GRACE}\hspace*{-0.5mm}     
&{\small \hspace*{-0.5mm}WOPPER\hspace*{-0.5mm}}   
&{\small \hspace*{-0.5mm}HERWIG}\hspace*{-0.5mm} 
&{\small \hspace*{-0.5mm}HERWIG}\hspace*{-0.5mm}
&{\small \hspace*{-0.5mm}COMPHEP\hspace*{-0.5mm}} 
&{\small \hspace*{-0.5mm}PYTHIA}\hspace*{-0.5mm} \\ \hline
  $-0.9$ & 6.86     &  14.9    &  14.4  & 0.771  & 1.17   &  ---  \\
  0 &           5.59      &  7.86    & 12.1   & 0.574  & 0.864     &  6.0   \\
  0.9 & 4.61      &  0.906   & 9.66   & 0.376  & 0.554        & ---  \\ \hline
\end{tabular}
\end{center}
\vspace*{-0.6cm}
\end{table}

\section{Results}
\begin{figure}[hp]
\vspace*{0.2cm}
\begin{minipage}{0.41\textwidth}
\caption{\label{fig:ellipse} Error bands and the corresponding error ellipse 
as a function of \mt\ and \cost\ for $\sqrt s =500$~GeV
and ${\cal L}=500$~fb$^{-1}$. 
The dot corresponds to $\mt=180$~GeV and $\cost=0.57$.
The errors from the 10~fb$^{-1}$ Morioka study were
$\Delta\mt$\,$=$\,7~GeV and $\Delta\cost$\,$=$\,0.06, 
while the 500~fb$^{-1}$ Sitges study
gives $\Delta\mt=1$~GeV and $\Delta\cost=0.009$.}
\end{minipage}
\hfill
\begin{minipage}{0.58\textwidth}
\vspace*{-0.3cm}
\begin{center}
\mbox{\epsfig{file=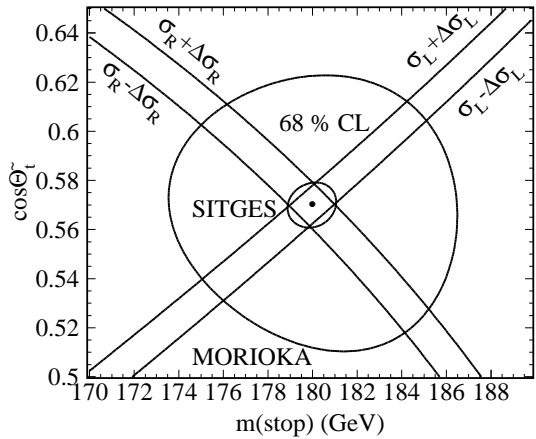,width=1.0\textwidth}}
\end{center}
\vspace*{-1cm}
\end{minipage}
\end{figure}

We have determined the expected background rate for a given signal
efficiency and checked that the discovery sensitivity for a 180 GeV scalar top
is almost independent of the working point signal efficiency in the range 5\% to 20\%. 
The total simulated background of about 16 million events is reduced to a few hundred, 
which allows a precision measurement of the scalar top production cross section with 
a relative error of better than 2\%.
Figure~\ref{fig:ellipse} shows the corresponding error bands and the error
ellipse in the \mt\ -- \cost\ plane.
The errors are a factor of 7 better than reported 
previously~\cite{zphys}.
In conclusion, an IDA analysis based on experience at LEP2
was applied and it  improved significantly the signal sensitivity. 
A high-luminosity linear collider with the capability of beam 
polarization has a great potential for precision measurements 
in the scalar quark sector predicted by supersymmetric theories.

\end{document}